\documentclass[a4paper]{jpconf}
\usepackage{graphicx}

\begin{document}
\title{Towards solar activity maximum 24 as seen by GOLF and VIRGO/SPM instruments}

\author{R. A. Garc\'\i a$^1$, D. Salabert$^{2}$, S. Mathur$^{3,4}$, C. R\'egulo$^{5,6}$, J. Ballot$^{7,8}$, G.R. Davies$^{1,9}$, A. Jim\'enez$^{5,6}$, R. Simoniello$^1$}
\address{$^1$ Laboratoire AIM, CEA/DSM-CNRS-Universit\'e Paris Diderot; CEA, IRFU, SAp, F-91191, Gif-sur-Yvette, France}
\address{$^2$ Laboratoire Lagrange, UMR7293, Universit\'e de Nice Sophia-Antipolis, CNRS, Observatoire de la C\^ote d'Azur, Bd. de l'Observatoire, 06304 Nice, France}
\address{$^3$ High Altitude Observatory, 3080 Center Green Drive, Boulder, CO, 80302 USA}
\address{$^4$ Space Science Institute, 4750 Walnut Street, Suite 205, Boulder, Colorado 80301 USA}
\address{$^5$ Instituto de Astrof\'isica de Canarias, E-38200 La Laguna, Tenerife, Spain}
\address{$^6$ Dept. de Astrof\'isica, Universidad de La Laguna, E-38206 La Laguna, Tenerife, Spain}
\address{$^7$  CNRS, Institut de Recherche en Astrophysique et Plan\'etologie, 14 avenue Edouard Belin, 31400 Toulouse, France}
\address{$^8$ Universit\'e de Toulouse, UPS-OMP, IRAP, 31400 Toulouse, France}
\address{$^9$ School of Physics and Astronomy, University of Birmingham, Edgbaston, Birmingham, B15 2TT, UK}

\ead{rgarcia@cea.fr, salabert@oca.eu, smathur@SpaceScience.org, crr@iac.es, Jerome.ballot@irap.omp.eu, davies@bison.ph.bham.ac.uk, ajm@iac.es, Rosaria.Simoniello@cea.fr}

\begin{abstract}
All p-mode parameters vary with time as a response to the changes induced by the cyclic behavior of solar magnetic activity. After the unusual long solar-activity minimum between cycles 23 and 24 --where the p-mode parameters have shown a different behavior than the surface magnetic proxies-- we analyze the temporal variation of low-degree p-mode parameters measured by GOLF (in velocity) and VIRGO (in intensity) Sun-as-a-star instruments on board SoHO.  We then compared our results with other activity proxies. 
\end{abstract}

\section{Introduction}
GOLF and VIRGO/SPM instruments have been used since 1996 to study the characteristics of low-degree, solar-acoustic modes \cite{Tou97} and even to unveil the asymptotic properties of dipole gravity modes \cite{Gar07,Gar08}. Thanks to their high-quality observations, we were able to measure all the p-mode properties with very high precision not previously attainable, including mode asymmetries \cite{Thi00}. Moreover, the temporal variations of the p-mode parameters during cycle 23 were studied \cite{Gel02,Jim03,Jim07}. The unexpected long activity minimum between cycles 23 and 24 \cite{Sal09} reminds us that the physical processes governing the magnetic activity in the Sun are not yet well understood.

\section{Observations and Data Analysis}
We analyzed observations collected by the space-based GOLF \cite{Gab95} and VIRGO \cite{Fro95} instruments onboard SoHO. A total of 6000 days were analyzed covering nearly 16.5 years between 1996 and 2012. These datasets were split into contiguous 365-day subseries, with a one-fourth overlap. The power spectrum of each subseries was fitted to extract the mode parameters using a standard likelihood maximization function (power spectrum with a $\chi^2$ with 2 d.o.f. statistics). Each mode component was parameterized using an asymmetric Lorentzian profile \cite{Nig98}. The temporal variations of the frequency shifts were defined as the difference between reference values (taken as the average over 1996-1997) and the parameters of the corresponding modes observed at different dates. Subseries with duty cycles less than 90\% (around the SoHO vacation) were not taken into account for this analysis. The weighted averages over the central part of the 5-min oscillation power --from 2200 to  3400 $\mu$Hz-- of the temporal variations of the mode parameters were then calculated. Mean values of daily measurements of the 10.7-cm radio flux were used as a proxy of the solar surface activity. Linear regressions were performed between the temporal variations of the mode parameters and the radio flux using independent points only.

\section{Frequency Variations of Individual Low-Degree p Modes}
\begin{figure}[!htb]
\includegraphics[width=27pc]{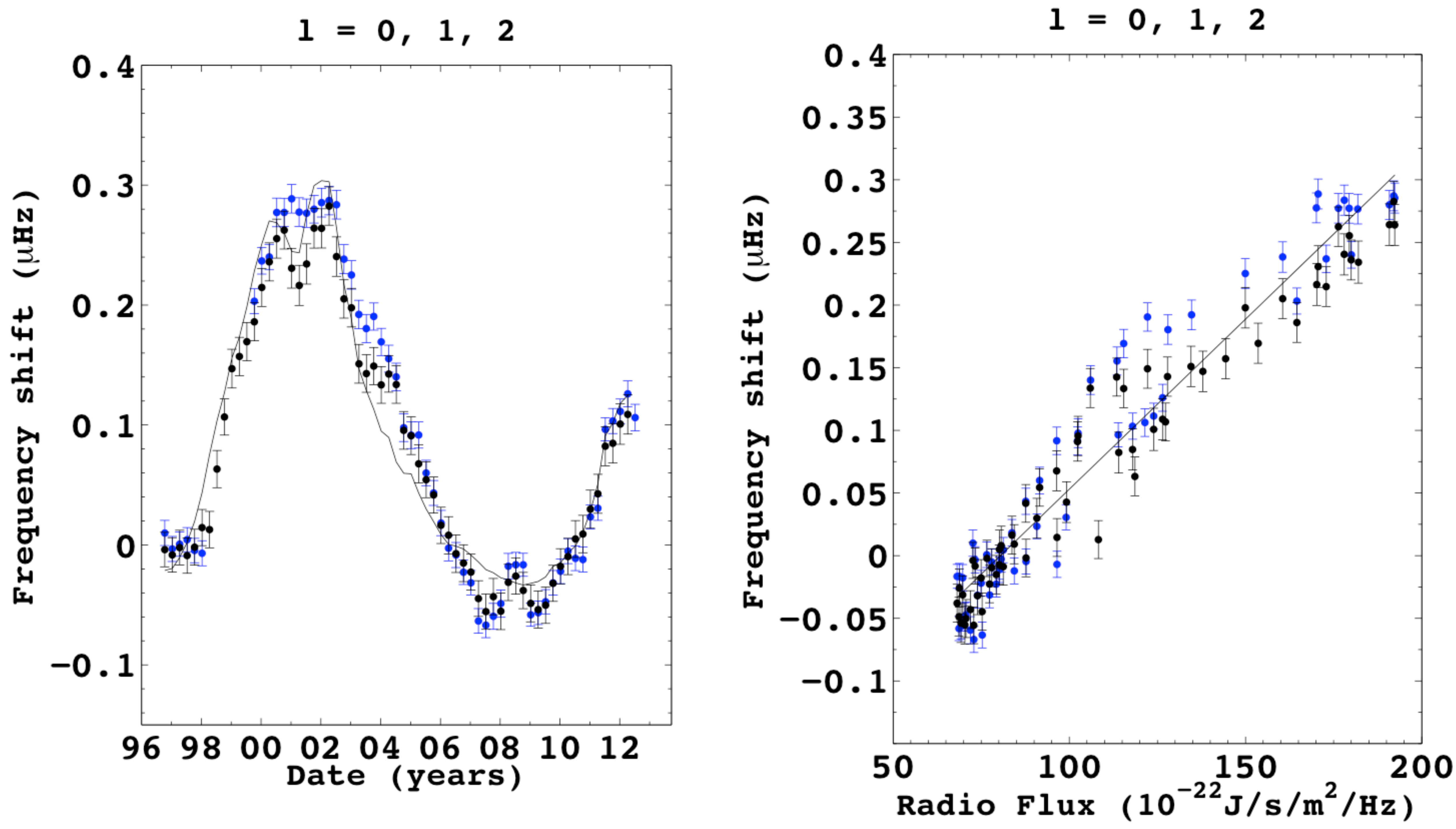}\hspace{1pc}%
\begin{minipage}[b]{10pc}\caption{\label{Fig1}Left panel: Temporal variations of  low-degree modes as described in the text; GOLF(black) and VIRGO (blue). The solid lines correspond to the scaled radio flux. Right panel: Frequency shifts as a function of the radio flux at 10.7 cm.}
\end{minipage}
\end{figure}

In Fig.~\ref{Fig1} we show the average temporal variations of the l=0,1, and 2 p-mode frequencies observed by GOLF and VIRGO following \cite{Sal08}. The 11-year solar cycle is clearly visible superimposed to the the 2-year modulation originally described by \cite{Bro09b}, and fully discussed by \cite{Fle10,Sim12}. The frequency shifts are anticorrelated with the amplitude of the modes. No significant temporal variations have been found on the rotational splittings in agreement with \cite{Bro12}.

\section{Mode Excitation and Damping with Solar Activity}
\begin{figure}[!htb]
\includegraphics[width = 0.7\textwidth]{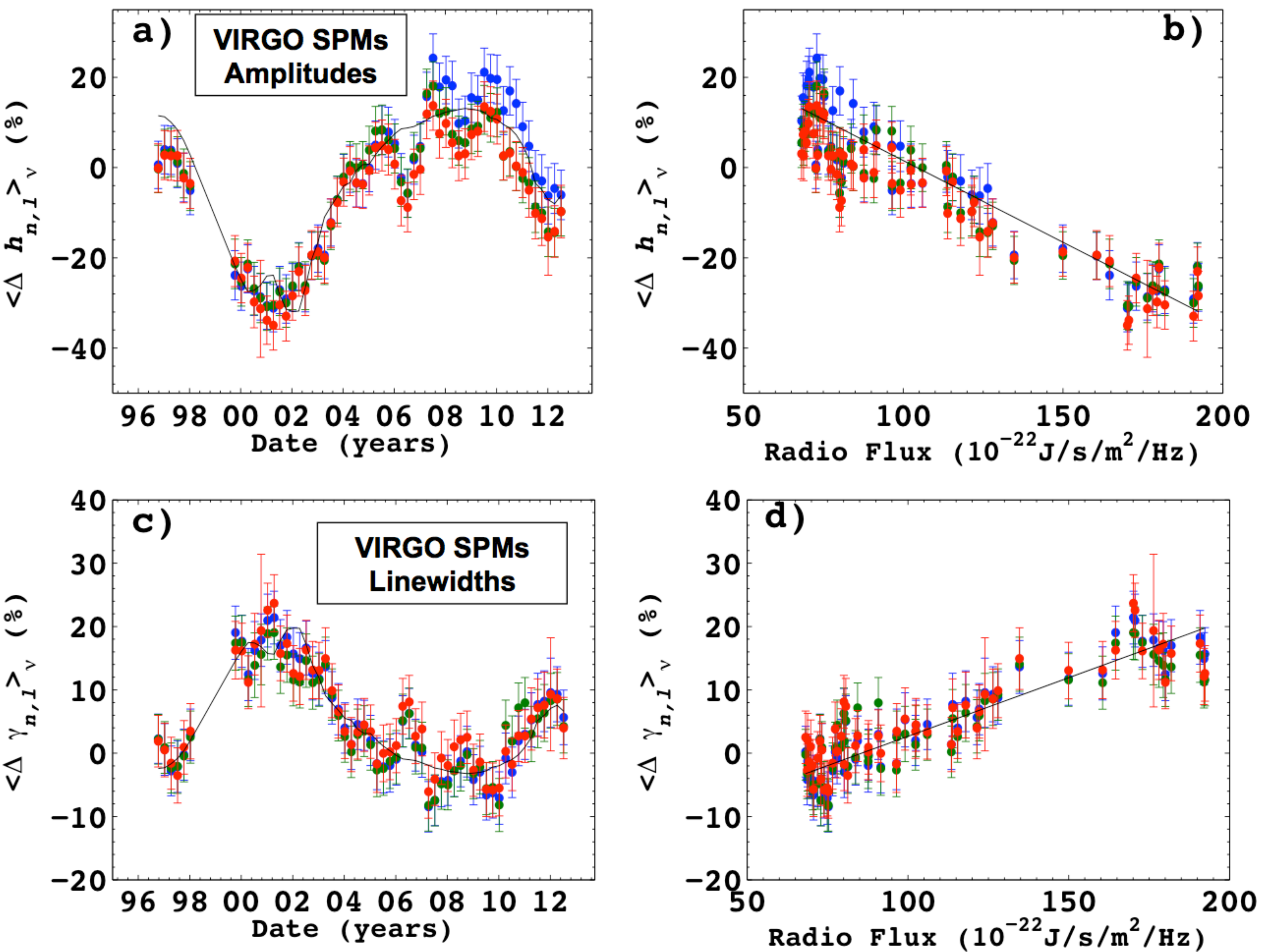}	
\begin{minipage}[b]{11pc}
\caption{\label{Fig2}Average of the amplitudes (a,b) and linewidths (c,d) of the modes  l = 0, 1 and 2 obtained using the three independent VIRGO channels: blue, green, and red as a function of time (left panels) and the radio flux (right panels). The solid lines correspond to the scaled radio flux.}
\end{minipage}
\end{figure}

The temporal variations of the mode amplitudes, $<$$\Delta h$$>$, and linewidths, $<$$\Delta \gamma$$>$,  are shown in Fig.~\ref{Fig2} computed using  the three VIRGO/SPM channels. Note that due to absolute calibration problems and the changes of the observing wings \cite{Gar05}, the GOLF amplitudes and linewidths are not exploitable for the moment for this analysis. A proper calibration is currently underway.   

\begin{figure}[!htb]
\includegraphics[width=18pc]{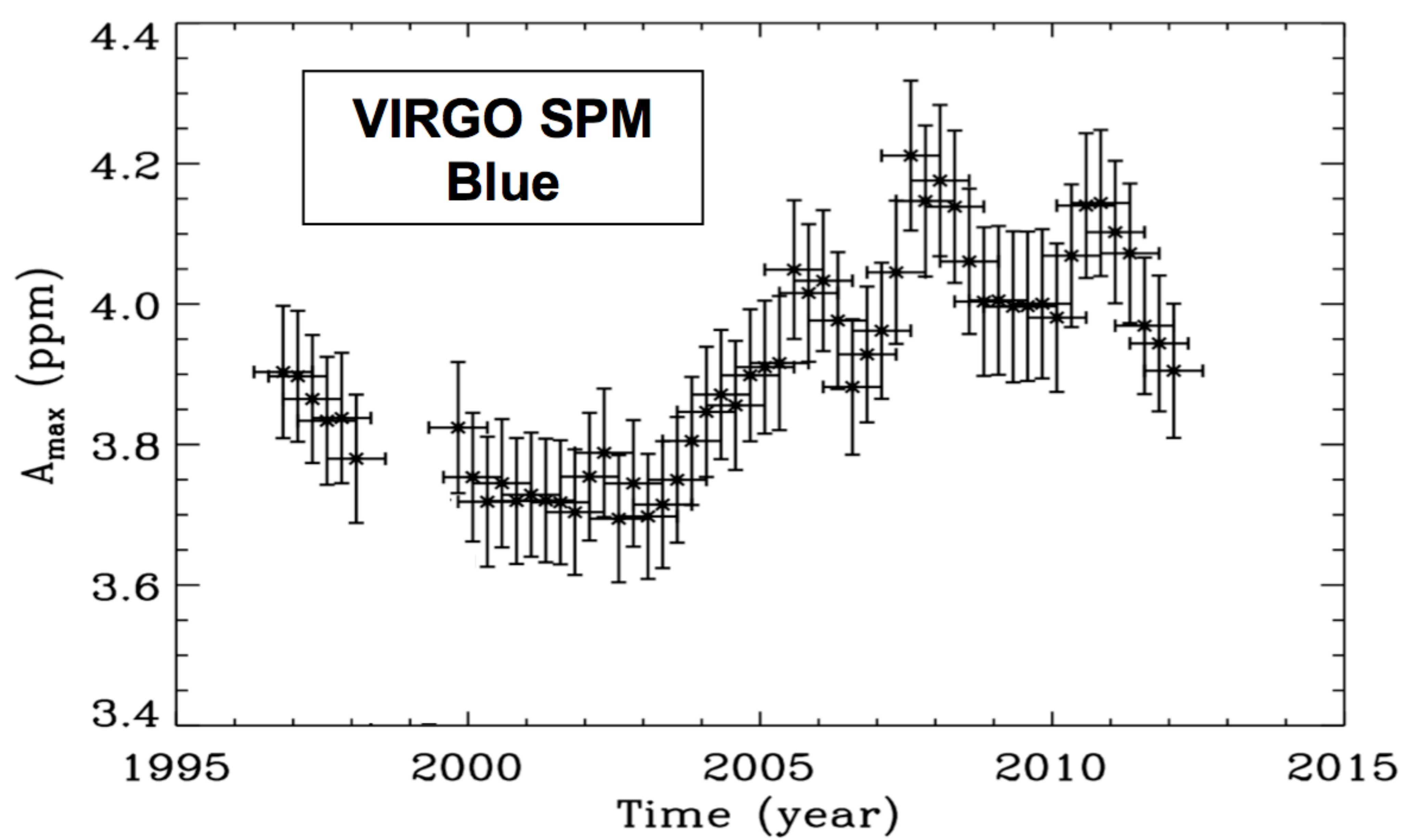}\hspace{1pc}%
\begin{minipage}[b]{19pc}\caption{\label{Fig3} Temporal variation of $A_{max}$ using the averaged three VIRGO/SPM channels as explained in the text.}
\end{minipage}
\end{figure}

In Fig.~\ref{Fig3}, we show the variations of $A_{max}$ \cite{Kje08} as it is usually done in asteroseismology to track down  activity cycles \cite{Gar10} computed following \cite{Mat10}. The results are comparable to the average of the individual mode amplitudes as shown in Fig.\ref{Fig2}a.

\section{Variation of Asymmetry} 
\begin{figure}[!htb]
\includegraphics[width = 0.6\textwidth]{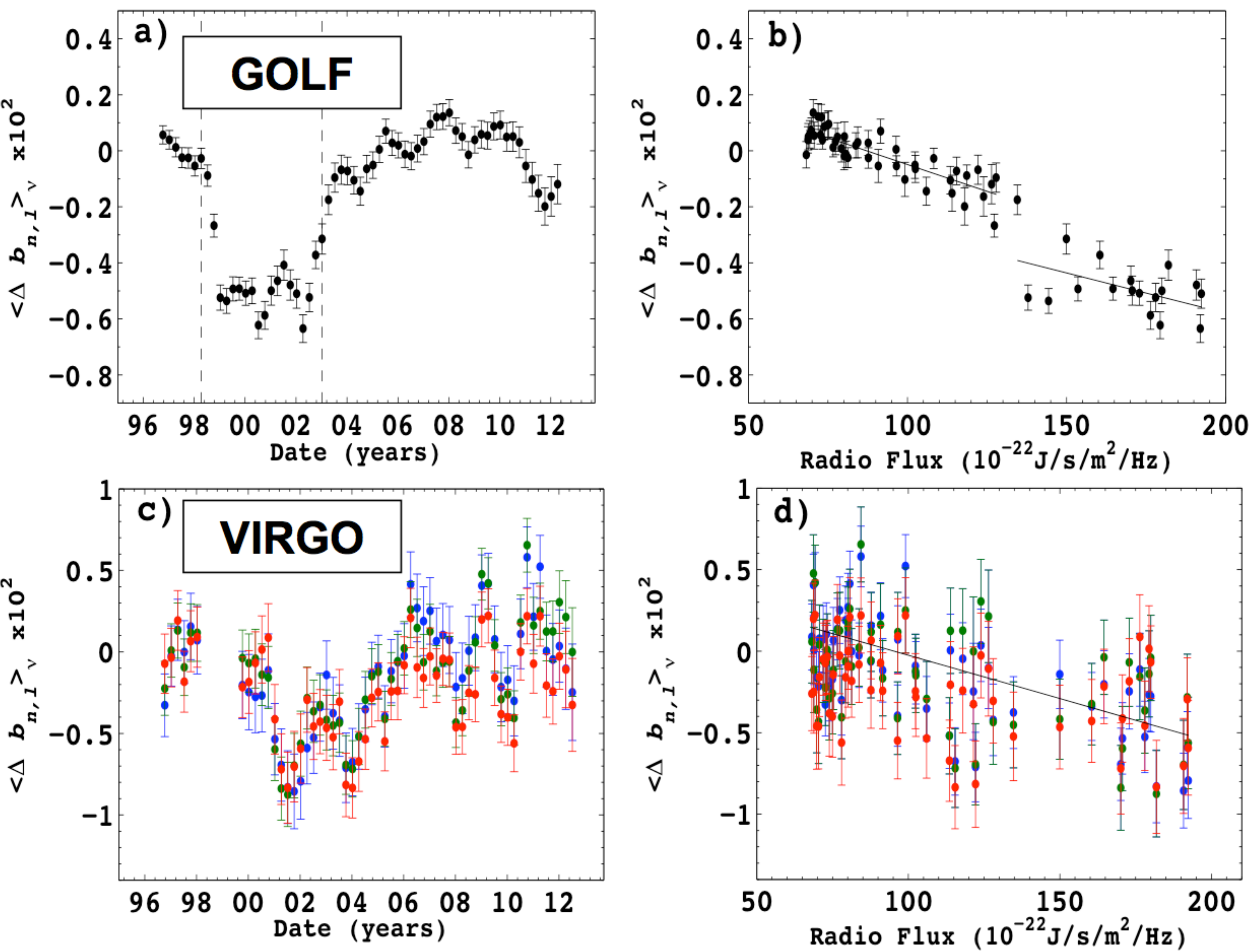}	
\begin{minipage}[b]{11pc}
\caption{\label{Fig4}Average of the mode asymmetries computed using GOLF (a,b) and VIRGO/SPM (c,d) data as a function of time (left panels) and radio flux (right panels). The solid lines correspond to the scaled radio flux.}
\end{minipage}
\end{figure}

In Fig.~\ref{Fig4}, the temporal variations of the peak asymmetry $<$$\Delta b$$>$ of the modes observed by GOLF and VIRGO/SPM are shown. Due to the change in the GOLF observing configuration between the blue and the red wings \cite{Gar05}, the variations (gradient) with the cycle are different (see Fig.4b).

\section{Solar Activity Proxy with GOLF and VIRGO/SPM}
By correcting the raw VIRGO/SPM averaged data using the algorithms used to process {\it Kepler} data \cite{Gar11}, we are able to measure the time evolution of the rotation signature produced by the sunspots crossing the visible solar disk. The projection of the wavelet power spectrum \cite{Tor98} in the range 6 to 60 days onto the time domain, provides us with a proxy of the 11-year magnetic activity cycle.  This methodology could be use to track magnetic activity cycles in other stars.

\begin{figure}[!htb]
\includegraphics[width=16.5pc]{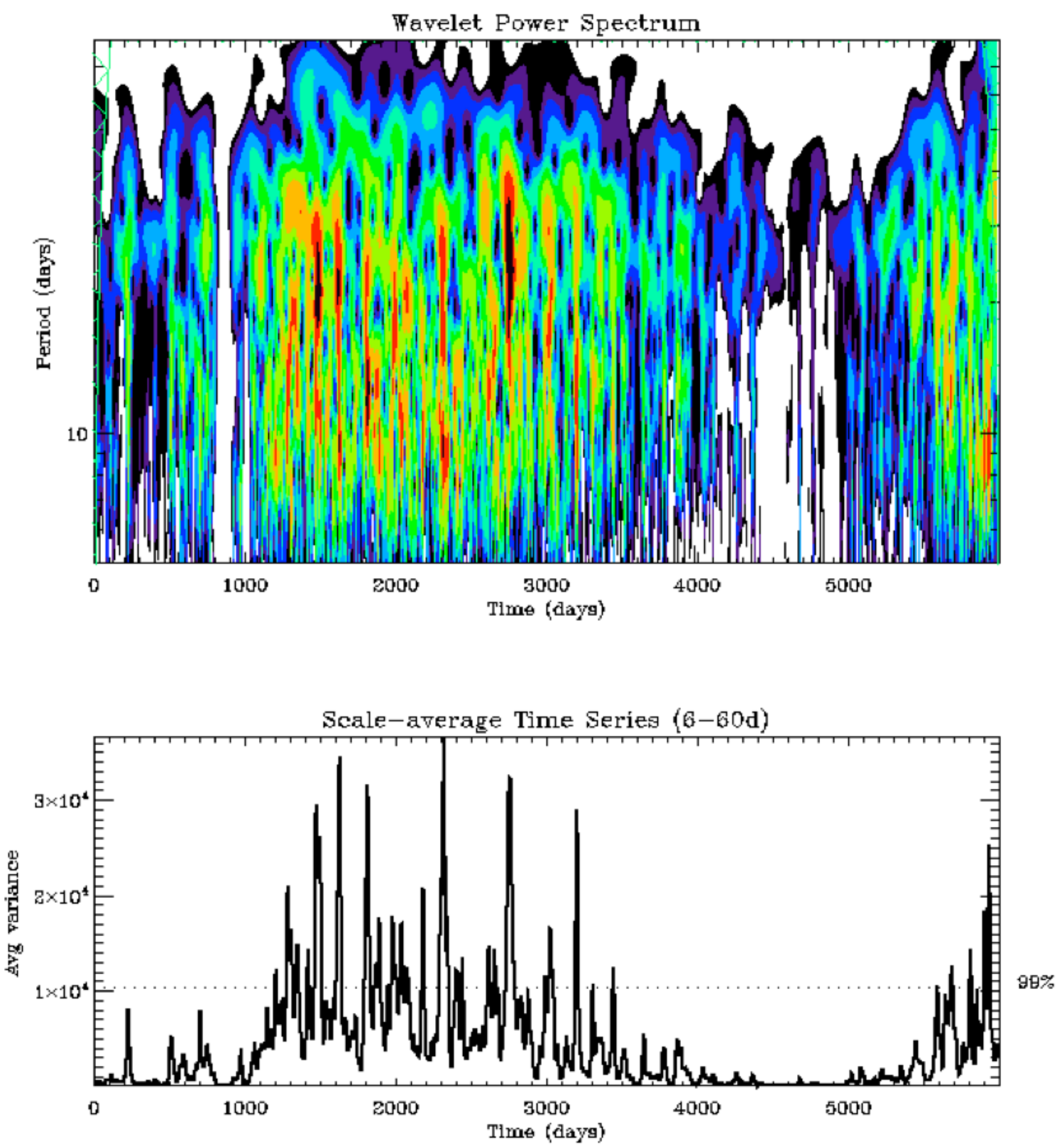}\hspace{1pc}%
\begin{minipage}[b]{19.5pc}\caption{\label{Fig5} Top: Wavelet power spectrum of VIRGO/SPM computed using \cite{Tor98, Mat10}. Bottom: Projection onto the time axis as explained in the text. the dotted line is the 99\% confident level. }
\end{minipage}
\end{figure}

\ack
SoHO is a space mission of international cooperation between ESA and NASA. R.A.G and D.S. thank the support from CNES. This work was partially supported by the NASA grant NNX09AE59G, by the White Dwarf Research Corporation through the Pale Blue Dot project, and by the grant AYA2010-17803 from the Spanish National Research Plan. NCAR is supported by the National Science Foundation. 


\end{document}